\documentclass[9pt,twocolumn,twoside]{osajnl}%
\usepackage{amsmath}
\usepackage{graphicx}
\usepackage{breakurl}%
\usepackage{amsfonts}%
\usepackage{amssymb}
\journal{josab} 
\setboolean{shortarticle}{false}
\ifthenelse{\boolean{shortarticle}}{\colorlet{color2}{color2b}}{\colorlet{color2}{color2}} 
\makeatletter
\renewcommand{\p@subsection}{\thesection\,}
\makeatother

\begin{document}
\title
{Efficient simulation of the swept-waveform polarization dynamics in fiber spools and Fourier domain mode-locked (FDML) lasers}
\author[1,*]{Christian Jirauschek}
\author[2]{Robert Huber}
\affil
[1]{Institute for Nanoelectronics, Technical University of Munich, D-80333 Munich, Germany}
\affil[2]{Institut f\"{u}r Biomedizinische Optik, Universit\"{a}%
t zu L\"{u}beck, D-23562 L\"{u}beck, Germany}
\affil[*]{Corresponding author: jirauschek@tum.de}
\dates{Journal reference: J. Opt. Soc. Am. B \bf{34}, 1135--1146 (2017)}
\ociscodes
{(140.3600) Lasers, tunable; (140.3430) Laser theory; (260.5430) Polarization; (260.1440) Birefringence; (320.5550) Pulses; (170.4500) Optical coherence tomography.}
\doi{\url{https://doi.org/10.1364/JOSAB.34.001135}}
\begin{abstract}
We present a theoretical model and its efficient numerical implementation for the simulation of wavelength-swept waveform propagation in fiber systems such as Fourier domain mode-locked (FDML) lasers, fully accounting for the polarization dynamics in fiber spools and further polarization dependent optical components in the setup. This approach enables us to perform long-time simulations of the FDML laser dynamics over more than 100000 cavity roundtrips, as required for some FDML configurations to ensure convergence to the steady state operating regime. The model is validated against experimental results for single propagation through a fiber spool and for stationary FDML operation. The polarization dynamics due to the fiber spool, inducing polarization-mode dispersion, bending birefringence as well as cross-phase modulation, and other optical components such as the Faraday-rotating mirror used for polarization compensation is thoroughly investigated.
\end{abstract}
\setboolean{displaycopyright}{true}%

\maketitle\thispagestyle{fancy}
\ifthenelse{\boolean{shortarticle}}{\ifthenelse{\boolean{singlecolumn}%
}{\abscontentformatted}{\abscontent}}{}

\section{Introduction}

For various applications such as optical coherence tomography (OCT)
\cite{huang1991optical}, broadband light sources with excellent coherence
properties are needed. These seemingly contradicting requirements can be
fulfilled by swept laser sources, performing fast narrowband frequency sweeps
over a wide spectral range. However, conventional tunable laser sources are
inherently limited in their sweep speed due to the required buildup time of
lasing in the cavity \cite{jirauschek2015wavelength}. In 2006, Fourier domain
mode locking (FDML) was discovered as a new stationary operating regime, which
circumvents this limitation and thus enables very high sweep rates
\cite{huber2006fourier}. The wavelength sweeps are here introduced by driving
an intracavity bandpass filter synchronously to the roundtrip time of the
circulating optical field in the laser cavity. To obtain a sufficiently large
roundtrip time adapted to the tuning rate of the bandpass filter, FDML lasers
require cavity lengths of up to a few km, which is achieved by using a fiber
ring resonator.

For FDML\ lasers, a sweep range of $284\,\mathrm{nm}$ has been demonstrated
using two semiconductor optical amplifiers (SOAs) arranged in a parallel
configuration \cite{hsu2011broadband}. Employing the so-called buffering
technique where the sweep rate is increased by combining time-delayed copies
of a wavelength sweep \cite{huber2006buffered}, sweep rates of up to
$5.2\,\mathrm{MHz}$ have been achieved
\cite{wieser2010multi,klein2011megahertz}. Furthermore, output powers of
around $100\,\mathrm{mW}$ have been obtained \cite{mao2009high}. Typical
instantaneous linewidths are below $0.1\,\mathrm{nm}$, corresponding to a
coherence length of up to centimeters \cite{adler2011extended}. This unique
combination of properties makes the FDML laser the system of choice for many
high speed OCT and sensing applications.

FDML-based OCT systems have been employed for various applications such as
noninvasive imaging of the human retina for identifying pathologies
\cite{klein2011megahertz}, and their potential for monitoring of therapy as
well as early stage cancer detection with OCT-based endomicroscopy has been
demonstrated \cite{adler2007three}. Furthermore, FDML sources are very
attractive for high speed sensing applications. In particular, FDML-based
fiber Bragg grating sensor systems, as used for temperature, pressure,
position and vibration measurements, have shown superior performance
\cite{Jung:08,lee2011remote,lee2010dynamic}. A very recent FDML-based
application is high speed stimulated Raman scattering spectroscopy and
microscopy, enabling quantitative chemical analysis of unknown samples and
label-free biomedical imaging, respectively \cite{karpf2015time}. Furthermore,
the temporal compressibility of the wavelength-swept FDML output to optical
pulses with durations of a few ten ps has been demonstrated, offering a new
approach to ultrashort optical pulse generation
\cite{eigenwillig2013picosecond}.

If no polarization maintaining (PM) optical components such as special PM
fiber are used, polarization effects greatly influence the FDML operating
characteristics due to effects such as cross-phase modulation, bending
birefringence and polarization-mode dispersion (PMD) in the fiber spool
\cite{wieser2012chromatic,jirauschek2015modeling}. Theoretical models for
FDML\ lasers typically do not take into account the polarization degree of
freedom of the optical field
\cite{jirauschek2009theoretical,slepneva2013dynamics,tu2012theoretical,li2013wkb,jirauschek2015wavelength}%
, and are thus only suitable for PM FDML setups. In this case, polarization is
controlled by using PM optical fiber and other components along with spliced
fiber connections to ensure accurate orientation of the optical axes
\cite{wieser2012chromatic}. Consequently, linearly polarized output is
generated, which makes PM FDML lasers suitable for use in polarization
sensitive OCT applications
\cite{bonesi2012high,bonesi2013high,INSPEC:10812530}. On the other hand, the
need for PM components and spliced connections renders the setup more
expensive. For this reason, commonly non-PM FDML configurations are used
\cite{wieser2010multi,adler2007three,klein2011megahertz,eigenwillig2013picosecond}%
, which are cheaper and easier to build. In these configurations, the
polarization state of the optical field changes in an uncontrolled manner,
which can be suppressed by using a polarization controller and a
single-polarization SOA \cite{wieser2012chromatic}.

In \cite{jirauschek2015modeling}, our well-established model equations
for\ PM\ FDML\ lasers
\cite{jirauschek2009theoretical,todor2011instantaneous,todor2012balance} have
been extended to account for the polarization dynamics in the fiber spool and
the other non-PM laser components. The large number of several million
numerical grid points necessary for sufficient temporal and spectral
resolution, combined with the necessity to simulate over $10^{4}-10^{5}$
roundtrips to obtain convergence, requires a highly efficient numerical
approach. Here, we give a detailed discussion of our polarization sensitive
FDML model, and present a computationally efficient implementation. In
comparison with \cite{jirauschek2015modeling} the model has been considerably
refined, among other things, to include a significantly enhanced
implementation of PMD with full frequency\ dependence. The obtained
theoretical approach is used to analyze polarization effects in a fiber spool
and in an FDML laser setup.

The paper is organized as follows: In Section \ref{sec:setup}, the setup of a
typical non-PM FDML laser is discussed. Section \ref{sec:theo} introduces the
theoretical model for non-PM FDML lasers, and in Section \ref{sec:num}, an
efficient numerical implementation is developed. In Section \ref{sec:results},
the model is validated against experimental results for single propagation
through a fiber spool and for stationary FDML operation. Furthermore, the
influence of the individual polarization effects is theoretically assessed.
The paper is concluded in Section \ref{sec:conclusion}.

\section{Non-Polarization Maintaining FDML Laser Setup}

\label{sec:setup}The most widely used type of FDML setup is a non-PM
configuration with a broadband single-polarization SOA, which typically
exhibits high gain and is available for a wide range of wavelength
specifications. Here, the outcoupled light shows good linear polarization
similar to the PM FDML design \cite{wieser2012chromatic}, but the polarization
state has to be adjusted using a polarization controller to obtain good
performance over the whole sweep range. Conventional FDML-based OCT
applications typically use this type of FDML setup
\cite{wieser2010multi,adler2007three,klein2011megahertz}, and also
polarization sensitive OCT has already been demonstrated with such a design
\cite{marschall2012high}.

Non-PM setups with polarization insensitive SOA can be used for applications
which are not polarization dependent. In contrast to non-PM setups with
single-polarization SOAs, no adjustment of the polarization controller is
required to obtain good performance. However, the polarization state
significantly changes over the sweep and furthermore exhibits large
sweep-to-sweep fluctuations \cite{wieser2012chromatic}. Moreover, polarization
insensitive broadband SOAs tend to have a lower gain than their polarization
dependent counterparts, and for certain wavelength ranges they are not
available. On the other hand, preliminary results indicate that non-PM FDML
configurations with polarization insensitive SOA have a smaller linewidth,
which is tentatively attributed to self-regulation effects using the
polarization degree of freedom \cite{jirauschek2015modeling}. Recently, the
excellent coherence properties of such setups have even been exploited to
generate picosecond pulses by external compression of the wavelength sweeps
\cite{eigenwillig2013picosecond}. The potential to control the pulse shape by
adapting the sweep filter drive waveform and the SOA current, as well as
optical pulse energies in the $\mu\mathrm{J}$ range, make this concept
appealing. \begin{figure}[ptbh]
\centerline{\includegraphics[width=8.4cm]{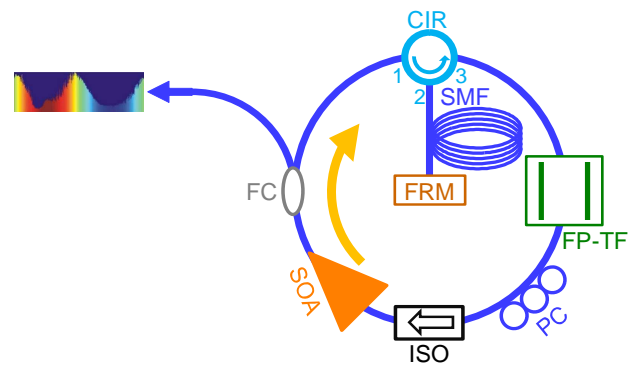}}\caption{Schematic
illustration of a non-polarization maintaining FDML laser setup.}%
\label{fig1}%
\end{figure}In Fig.\thinspace\ref{fig1}, a non-PM FDML laser setup is
schematically illustrated. As a gain medium, an SOA is used, followed by a
fiber output coupler (FC) where part of the light is outcoupled. To ensure
unidirectional lasing, an optical isolator (ISO) is built in before the SOA.
Using a circulator (CIR), the light is coupled into a single-mode fiber (SMF)
spool serving as a delay line. The effective cavity length is doubled by
back-reflecting the light at the fiber end. For this purpose, a Faraday
rotating mirror (FRM) is used which also rotates the polarization angle by 90%
${{}^\circ}$
to achieve a partial compensation of polarization effects in the optical
fiber. A Fabry-P\'{e}rot tunable filter (FP-TF) with a bandwidth of typically
$\sim0.15\,\mathrm{nm}$ serves for wavelength selection, while the
instantaneous linewidth of the laser light is considerably narrower. Rapid
wavelength tuning is here achieved by synchronizing the tuning rate of the
FP-TF to the optical roundtrip time of light circulating in the laser cavity,
which avoids repeated build-up of the optical resonator field.

\section{Theoretical Model}

\label{sec:theo}

Important fiber propagation effects are self-phase modulation (SPM) due to an
intensity dependent refractive index and dispersion due to the dependence of
the propagation constant $\beta$ on the (angular) frequency $\omega$.
Commonly, $\beta$ is represented by its Taylor series about the center
frequency $\omega_{\mathrm{c}}$ of the optical field,
\begin{equation}
\beta\left(  \omega\right)  =\sum_{j}\left(  \omega-\omega_{\mathrm{c}%
}\right)  ^{j}\beta^{\left(  j\right)  }/j!, \label{beta}%
\end{equation}
with $\beta^{\left(  j\right)  }=\partial_{\omega}^{j}\beta\left(
\omega_{\mathrm{c}}\right)  $. The polarization effects in the optical fiber
are mainly related to birefringence, causing different phase and group
velocities for the electric field components $E_{x,y}$ in the two polarization
directions $x$ and $y$, i.e., $\beta_{x}^{\left(  0\right)  }\neq\beta
_{y}^{\left(  0\right)  }$ and $\beta_{x}^{\left(  1\right)  }\neq\beta
_{y}^{\left(  1\right)  }$. Bending birefringence is due to lateral internal
stress in bent optical fibers, such as fibers wound on a spool or fiber loops
used for polarization control. Besides, imperfections such as anisotropic
stresses and deviations from rotational symmetry lead to additional
birefringence contributions which vary along the fiber, giving rise to PMD.
The modeling of this effect poses an extra challenge due to the random nature
of the imperfections, requiring a stochastic treatment. The nonlinear
birefringence contribution accompanying SPM leads to coupling between the two
field polarization components, which is referred to as cross-phase modulation (XPM).

We describe the optical field by its slowly varying envelope functions
$A_{x,y}\left(  z,\tau\right)  $, defined so that $E_{x,y}\propto\Re\left\{
A_{x,y}\exp\left[  \mathrm{i}\left(  \beta_{x}^{\left(  0\right)  }+\beta
_{y}^{\left(  0\right)  }\right)  z/2-\mathrm{i}\omega_{\mathrm{c}}%
\tau\right]  \right\}  $ and the optical power is $\left|  A_{x}\right|
^{2}+\left|  A_{y}\right|  ^{2}$, where $z$ is the propagation coordinate in
the laser system and $\tau$ represents the time variable. Introducing the
retarded time $t=\tau-\left(  \beta_{x}^{\left(  1\right)  }+\beta
_{y}^{\left(  1\right)  }\right)  z/2$ defined with respect to a frame moving
along with the optical field, propagation through an optical fiber system can
be described by coupled nonlinear Schr\"{o}dinger equations for the two
polarization components \cite{agr01},%
\begin{align}
\partial_{z}A_{x,y}  &  =\left[  \pm\frac{\mathrm{i}}{2}\Delta_{\beta
,\mathrm{b}}\left(  \omega_{\mathrm{c}}+\mathrm{i}\partial_{t}\right)
+\mathrm{i}\sum_{j\geq2}D_{j}\left(  \mathrm{i}\partial_{t}\right)
^{j}\right]  A_{x,y}\nonumber\\
&  +\mathrm{i}\gamma\left(  \left|  A_{x,y}\right|  ^{2}+\frac{2}{3}\left|
A_{y,x}\right|  ^{2}\right)  A_{x,y}\nonumber\\
&  +\left[  g_{x,y}\left(  \omega_{\mathrm{c}}+\mathrm{i}\partial_{t}\right)
\left(  1-\mathrm{i}\alpha\right)  -a_{x,y}\left(  \omega_{\mathrm{c}%
}+\mathrm{i}\partial_{t}\right)  \right]  A_{x,y}\nonumber\\
&  +\frac{\mathrm{i}}{2}\Delta_{\beta,\mathrm{PMD}}\left(  \omega_{\mathrm{c}%
}+\mathrm{i}\partial_{t}\right)  \left[  \pm\cos\left(  \theta\right)
A_{x,y}+\sin\left(  \theta\right)  A_{y,x}\right]  . \label{nls}%
\end{align}
The first line contains birefringence and dispersion effects, where
$\Delta_{\beta,\mathrm{b}}=\beta_{x}-\beta_{y}$ denotes the difference of the
propagation constants for the two polarization directions due to bending
birefringence, see Section \ref{sec:BB}. Since $\beta$ is a function of
$\omega$, see Eq.\thinspace(\ref{beta}), also $\Delta_{\beta,\mathrm{b}}$ is
frequency dependent, which is in time domain reflected by the operator
function $\Delta_{\beta,\mathrm{b}}\left(  \omega_{\mathrm{c}}+\mathrm{i}%
\partial_{t}\right)  $. The second line contains SPM and XPM. Chromatic
dispersion, characterized by $j$th order dispersion coefficients $D_{j}=$
$\beta^{\left(  j\right)  }/j!$ with $j=2,3,\dots$, as well as the nonlinear
coefficient $\gamma$ describing SPM and XPM, is assumed to be identical for
both polarization directions \cite{agr01}. The third line describes frequency
dependent amplitude gain and loss, represented by separate coefficients
$g_{x,y}\left(  \omega\right)  $ and $a_{x,y}\left(  \omega\right)  $ for the
two polarization directions to account for the polarization sensitivity of
optical components such as the SOA module. Linewidth enhancement of the SOA is
taken into account by the Henry factor $\alpha$, and gain saturation is
modeled by introducing an optical power dependence into $g_{x,y}$
\cite{jirauschek2009theoretical,todor2012balance}. In the fourth line, PMD in
the fiber is implemented according to
\cite{wai1994polarization,1996JLwT...14..148W}. In this approach, the random
nature of PMD is reflected by fixed PMD-related birefringence contributions
$\Delta_{\beta,\mathrm{PMD}}$ at randomly varying orientation angles $\theta$
with $\partial_{z}\theta=g_{\theta}\left(  z\right)  $. Here, $g_{\theta
}\left(  z\right)  $ is a white noise process with $\left\langle g_{\theta
}\left(  z\right)  \right\rangle =0$, $\left\langle g_{\theta}\left(
z\right)  g_{\theta}\left(  z^{\prime}\right)  \right\rangle =2h_{\mathrm{f}%
}^{-1}\delta\left(  z-z^{\prime}\right)  $, and $h_{\mathrm{f}}$ denotes the
fiber autocorrelation length \cite{wai1994polarization,1996JLwT...14..148W}.

The number of required simulation grid points for sufficient temporal and
spectral resolution is set by the time-bandwidth product of the field
components $A_{x,y}$, which in FDML lasers corresponds to the product of the
sweep bandwidth and cavity roundtrip time, typically $\sim10\,\mathrm{THz}%
\times10\,\mu\mathrm{s}=10^{8}$. This impedes a direct numerical evaluation of
Eq.\thinspace(\ref{nls}). The computational load can be considerably reduced
by introducing a sliding spectral reference frame moving along with the time
dependent center frequency $\Omega\left(  t\right)  $ of the sweep filter, and
transforming $A_{x,y}$ to the swept filter reference frame
\cite{jirauschek2009theoretical}, with%
\begin{equation}
u_{x,y}\left(  z,t\right)  =A_{x,y}\left(  z,t\right)  \exp\left\{
\mathrm{i}\int^{t}\left[  \Omega\left(  t^{\prime}\right)  -\omega
_{\mathrm{c}}\right]  \mathrm{d}t^{\prime}\right\}  . \label{u}%
\end{equation}
Inserting Eq.\thinspace(\ref{u}) into Eq.\thinspace(\ref{nls}) and
approximating $\partial_{t}^{j}A_{x,y}\approx\left[  -\mathrm{i}\left(
\Omega-\omega_{\mathrm{c}}\right)  \right]  ^{j}u_{x,y}$, i.e., neglecting
higher order terms \cite{jirauschek2009theoretical}, yields%
\begin{align}
\partial_{z}\mathbf{u}  &  =\left[  \mathrm{i}\sum_{j\geq2}D_{j}\left(
\Omega-\omega_{\mathrm{c}}\right)  ^{j}+\mathrm{i}\gamma\left(
\begin{array}
[c]{cc}%
\left|  u_{x}^{2}\right|  & 0\\
0 & \left|  u_{y}^{2}\right|
\end{array}
\right)  -a_{\mathrm{s}}\left(  \mathrm{i}\partial_{t}\right)  \right.
\nonumber\\
&  +\left(
\begin{array}
[c]{cc}%
g_{x}\left(  \Omega\right)  & 0\\
0 & g_{y}\left(  \Omega\right)
\end{array}
\right)  \left(  1-\mathrm{i}\alpha\right)  -\left(
\begin{array}
[c]{cc}%
a_{x}\left(  \Omega\right)  & 0\\
0 & a_{y}\left(  \Omega\right)
\end{array}
\right) \nonumber\\
&  +\frac{\mathrm{i}}{2}\Delta_{\beta,\mathrm{PMD}}\left(  \Omega\right)
\left(
\begin{array}
[c]{cc}%
\cos\theta & \sin\theta\\
\sin\theta & -\cos\theta
\end{array}
\right) \nonumber\\
&  \left.  +\mathrm{i}\frac{2}{3}\gamma\left(
\begin{array}
[c]{cc}%
\left|  u_{y}^{2}\right|  & 0\\
0 & \left|  u_{x}^{2}\right|
\end{array}
\right)  +\frac{\mathrm{i}}{2}\Delta_{\beta,\mathrm{b}}\left(  \Omega\right)
\left(
\begin{array}
[c]{cc}%
1 & 0\\
0 & -1
\end{array}
\right)  \right]  \mathbf{u}, \label{prop}%
\end{align}
where $\mathbf{u}=\left[  u_{x},u_{y}\right]  ^{\mathrm{T}}$ and $\mathrm{T}$
denotes the transpose. In this new reference frame, the sweep filter can
simply be described by its static bandpass characteristics, here represented
by a frequency dependent loss coefficient $a_{\mathrm{s}}\left(
\omega\right)  $ centered at $\omega=0$. In time domain, this corresponds to
an operator function $a_{\mathrm{s}}\left(  \mathrm{i}\partial_{t}\right)  $.
This implies that even for a rapidly wavelength-tuned filter acting on a
suitably wavelength-swept optical field, the field ``sees'' a stationary
bandpass filter in the swept filter reference frame, as has been verified by
an exact treatment based on Maxwell's equations
\cite{jirauschek2015wavelength}. The transformed spectrum is now limited by
the sweep filter bandwidth rather than the sweep range, reducing the number of
required grid points by about two to three orders of magnitude. Setting
$g_{x}=g_{y}=g$ and $a_{x}=a_{y}=a$, the first and second line of
Eq.\thinspace(\ref{prop}) correspond to the previously developed FDML model
without polarization effects
\cite{jirauschek2009theoretical,todor2011instantaneous,todor2012balance}. The
third and fourth line contain fiber polarization effects due to PMD, XPM and
bending birefringence. The used models for $\Delta_{\beta,\mathrm{b}}$\ and
$\Delta_{\beta,\mathrm{PMD}}$ are discussed in Sections \ref{sec:BB} and
\ref{sec:PMD}, respectively.\begin{figure}[ptbh]
\centerline{\includegraphics[width=8.4cm]{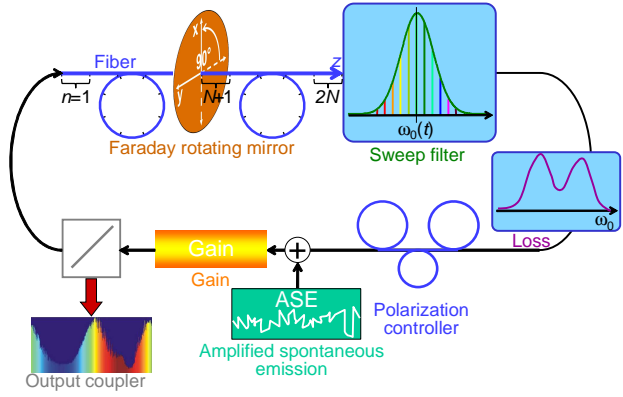}}\caption{Schematic diagram
of the FDML model used for the simulation of the laser setup sketched in Fig.
\ref{fig1}.}%
\label{fig2}%
\end{figure}

In Fig.\thinspace\ref{fig2}, the FDML model employed for simulating non-PM
laser setups as shown in Fig.\thinspace\ref{fig1} is schematically
illustrated. Here, the subsystem consisting of the circulator, fiber spool and
FRM is unfolded, taking advantage of the fact that the interaction of the
forward and backward propagating light in the fiber can be neglected. All the
parameters depend on the position $z$; for example, the coefficients $g_{x,y}$
are non-zero only inside the gain medium. A self-consistent laser model
requires the inclusion of saturation effects, as can be done by introducing
power dependent gain coefficients. Besides SOAs, for example also erbium-doped
and Raman fiber gain media have been employed for FDML\ lasers
\cite{lee2009fdml,klein2008raman}, and the details of the gain model depend on
the type of optical amplifier used. The sweep filter is modeled as a lumped
component, described by a complex Lorentzian amplitude transmission profile
$t_{\mathrm{s}}\left(  \omega\right)  =\exp\left[  -\int a_{\mathrm{s}}\left(
\omega\right)  \mathrm{d}z\right]  =T_{\max}^{1/2}/\left(  1-2\mathrm{i}%
\omega/\Delta_{\mathrm{s}}\right)  $ with the full width at half-maximum
(FWHM) bandwidth $\Delta_{\mathrm{s}}$ and peak transmittance $T_{\max}$
\cite{jirauschek2009theoretical,todor2012balance,jirauschek2015wavelength}. In
addition to the effects included in Eq.\thinspace(\ref{prop}), amplified
spontaneous emission (ASE) in the gain medium is considered by an equivalent
noise source at the input of the gain medium
\cite{cassioli2000time,jirauschek2009theoretical,todor2012balance}.
Furthermore, the Faraday rotating mirror at the end of the fiber spool is
modeled as discussed in Section \ref{sec:FRM}. The polarization controller
consists of three fiber loops which are twisted against each other. This
allows for an adjustment of the polarization state by exploiting bending
birefringence in the loops, already included in Eq.\thinspace(\ref{prop}). The
resulting model equations for the polarization controller are derived in
Section \ref{sec:PC}.

\subsection{\label{sec:BB}Bending Birefringence}

For a fiber with a cladding radius $r$ and a bending radius $R_{\mathrm{b}}$,
the bending-induced difference between the propagation constants of the two
axes is given by \cite{ulrich1980bending,namihira1985opto}%
\begin{align}
\Delta_{\beta,\mathrm{b}} &  =\beta_{x}-\beta_{y}=\omega c^{-1}n_{0}%
^{3}\left(  1+\nu\right)  \left(  p_{11}-p_{12}\right)  \left(
r/R_{\mathrm{b}}\right)  ^{2}/4\nonumber\\
&  =-\omega c^{-1}EC\left(  r/R_{\mathrm{b}}\right)  ^{2}/2,\label{Db}%
\end{align}
where $x$ indicates the component in the direction of the bending radius, and
$y$ is the component orthogonal to the coil plane. Furthermore, $\nu$ is
Poisson's ratio, $p_{11}$ and $p_{12}$ are elements of the photoelastic tensor
in silica, $n_{0}$ is the refractive index, and $c$ denotes the vacuum speed
of light. Young's modulus $E$ is a purely mechanical constant and thus
wavelength independent, with $E=72.7\,\mathrm{GPa}$ for fused silica
\cite{hurley2007measurement}. The dependence of the relative optoelastic
constant $C$ on the vacuum wavelength $\lambda=2\pi c/\omega$ is for
single-mode silica fibers given by the empirical equation $C\approx\sum
_{\ell=0}^{3}C_{\ell}\left(  \lambda/\mu\mathrm{m}\right)  ^{\ell}%
\times9.80665^{-1}\times10^{-5}\,\mathrm{mm}^{2}/\mathrm{N}$ with
$C_{0}=3.95065$, $C_{1}=-1.72552$, $C_{2}=1.52460$, $C_{3}=-0.50708$,
extracted from measurements of $C$ for wavelengths $0.6\,\mu\mathrm{m}%
<\lambda<1.6\,\mu\mathrm{m}$ \cite{namihira1985opto}. The factor
$9.80665^{-1}$ has been added to convert from the units $\mathrm{mm}%
^{2}/\mathrm{kg}$ used in \cite{namihira1985opto} to the units used here. With
Eq.$\,$(\ref{Db}), we obtain
\begin{align}
\Delta_{\beta,\mathrm{b}}\left(  \omega\right)   &  =\left[  -0.488\left(
\omega/\mathrm{fs}^{-1}\right)  +0.402-0.669\left(  \omega/\mathrm{fs}%
^{-1}\right)  ^{-1}\right.  \nonumber\\
&  \left.  +0.419\left(  \omega/\mathrm{fs}^{-1}\right)  ^{-2}\right]  \left(
r/R_{\mathrm{b}}\right)  ^{2}\times10^{6}\,\mathrm{m}^{-1}.\label{Dw}%
\end{align}
The optical fibers used in the fiber spool and FDML laser setup investigated
here have a cladding radius $r=62.5\,\mu\mathrm{m}$ \cite{SMF28}.

\subsection{\label{sec:PMD}Polarization-mode dispersion}

PMD in a fiber is usually characterized in terms of the PMD parameter
$D_{\mathrm{p}}$\ and fiber autocorrelation length $h_{\mathrm{f}}$ at the
frequency of interest $\omega_{\mathrm{c}}$ \cite{agr01}. From these
quantities, the first order derivative of $\Delta_{\beta,\mathrm{PMD}}$ is
obtained as \cite{agr01}
\begin{equation}
\frac{\mathrm{d}}{\mathrm{d}\omega}\Delta_{\beta,\mathrm{PMD}}=\Delta
_{\beta,\mathrm{PMD}}^{\prime}=D_{\mathrm{p}}\left(  2h_{\mathrm{f}}\right)
^{-1/2}.\label{DPMDs}%
\end{equation}
For our FDML model, Eq.$\,$(\ref{prop}), the knowledge of $\Delta
_{\beta,\mathrm{PMD}}^{\prime}$ and $h_{\mathrm{f}}$\ is not sufficient, but
rather the dependence of $\Delta_{\beta,\mathrm{PMD}}\ $on $\omega$ must be
known over the the full sweep bandwidth, which can be obtained from a suitable
physical model. PMD in single-mode optical fibers is due to imperfections
introduced during manufacturing, such as a noncircular core giving rise to
geometry-related and stress-related birefringence. The latter is caused by
non-symmetrical stress, which builds up during the cooling process due to
deviations from the circular fiber geometry, and dominates for fibers with
small index contrast \cite{sakai1982birefringence}. Using the frequency
dependent expression for $\Delta_{\beta,\mathrm{b}}\left(  \omega\right)  $
given in Eq.$\,$(\ref{Db}), the stress birefringence can be written as \cite{sakai1982birefringence}%

\begin{align}
\Delta_{\beta,\mathrm{PMD}}  &  =\varepsilon\frac{\left(  a_{1}-\alpha
_{2}\right)  \left(  T_{\mathrm{r}}-T_{\mathrm{s}}\right)  }{\left(
1-\nu\right)  }\left(  \frac{R_{\mathrm{b}}}{r}\right)  ^{2}\Delta
_{\beta,\mathrm{b}}\left(  \omega\right)  F\left(  \omega\right)  ,\nonumber\\
F\left(  \omega\right)   &  =\frac{W^{2}}{V^{2}}-3\frac{a_{\mathrm{c}}^{4}%
}{r^{4}}+\frac{a_{\mathrm{c}}^{6}}{r^{6}}\left[  2+\frac{4\left(  U^{2}%
-W^{2}\right)  }{U^{2}W^{2}}+\frac{4}{U}\frac{\mathrm{J}_{0}\left(  U\right)
}{\mathrm{J}_{1}\left(  U\right)  }\right]  . \label{DPMD}%
\end{align}
Here, $V=a_{\mathrm{c}}\omega c^{-1}n_{0}\left(  2\Delta\right)  ^{1/2}$ is
the normalized frequency, $W$ is approximately given by $W\approx
1.1428V-0.9960$ (with an error $<0.1\%$ for $1.5\leq V\leq2.5$)
\cite{rudolph1976approximations}, and $U=\left(  V^{2}-W^{2}\right)  ^{1/2}$.
The $\mathrm{J}_{n}$ are Bessel functions of the first kind. For an SMF-28
single-mode fiber, we have an (average) core radius of $a_{\mathrm{c}%
}=4.1\,\mu\mathrm{m}$ and a refractive index difference between core and
cladding of $\Delta=0.0036$ \cite{SMF28}. Furthermore, $\varepsilon$,
$a_{1,2}$, $T_{\mathrm{r},\mathrm{s}}$ and $\nu$ denote the core ellipticity,
thermal expansion coefficients of the core and cladding, room temperature and
glass softening temperature, and Poisson's ratio, respectively. It is
practical to rewrite Eq.$\,$(\ref{DPMD}) by using Eq.$\,$(\ref{DPMDs}), which
gives%
\begin{equation}
\Delta_{\beta,\mathrm{PMD}}=\Delta_{\beta,\mathrm{PMD}}^{\prime}\left(
\omega_{\mathrm{c}}\right)  \frac{\Delta_{\beta,\mathrm{b}}\left(
\omega\right)  F\left(  \omega\right)  }{\left.  \frac{\mathrm{d}}%
{\mathrm{d}\omega}\left[  \Delta_{\beta,\mathrm{b}}\left(  \omega\right)
F\left(  \omega\right)  \right]  \right|  _{\omega=\omega_{\mathrm{c}}}}.
\label{DPMD2}%
\end{equation}
\qquad For our simulations, we assume typical values of $D_{\mathrm{p}%
}=0.05\,\mathrm{ps}/\mathrm{km}^{1/2}$ and $h_{\mathrm{f}}=10\,\mathrm{m}$
\cite{agr01,SMF28}, yielding with Eq.$\,$(\ref{DPMDs}) $\Delta_{\beta
,\mathrm{PMD}}^{\prime}=0.354\,\mathrm{ps}/\mathrm{km}$.

\subsection{\label{sec:FRM}Faraday Rotating Mirror}

As shown in Fig.\thinspace\ref{fig1}, the Faraday rotating mirror (FRM)
back-reflects the light at the end of the fiber spool with a polarization
angle rotated by 90%
${{}^\circ}$%
. Ideally, this results in a compensation of polarization effects in the
spool, at least for low power levels where nonlinear optical effects due to
SPM\ and XPM are negligible. In practice, a 90%
${{}^\circ}$
rotation is only obtained at the FRM design wavelength $\lambda_{\mathrm{F}}$,
preventing a full compensation of the polarization effects over the complete
wavelength sweep range. Thus, we use a more realistic model of the FRM,
allowing us to capture these residual effects. The FRM utilizes the Faraday
effect where a magnetic field of strength $B$ in the optical propagation
direction causes a rotation of the polarization angle by $\phi=BV_{\mathrm{F}%
}L_{\mathrm{F}}$, with the Verdet constant of the material $V_{\mathrm{F}}$
and the total path length of the light $L_{\mathrm{F}}$
\cite{jenkins1976fundamentals}. For a paramagnetic material, $V_{\mathrm{F}}$
is expected to depend on the wavelength $\lambda$ according to $V_{\mathrm{F}%
}=\Xi/\left(  \lambda^{2}-\lambda_{\mathrm{r}}^{2}\right)  $, where
$\lambda_{\mathrm{r}}$ denotes the wavelength corresponding to the ultraviolet
resonance peak and $\Xi$ is a material constant \cite{barnes1992variation}.
For the FDML\ laser, the wavelength of the light field is determined by the
center wavelength of the sweep filter, $\lambda=2\pi c/\Omega$. Using $\phi=$
$B\Xi L_{\mathrm{F}}/\left(  \lambda_{\mathrm{F}}^{2}-\lambda_{\mathrm{r}}%
^{2}\right)  =\pi/2$ at the design wavelength $\lambda_{\mathrm{F}}$, we
obtain%
\begin{equation}
\phi\left(  \Omega\right)  =\frac{\pi}{2}\frac{\lambda_{\mathrm{F}}%
^{2}-\lambda_{\mathrm{r}}^{2}}{\left(  2\pi c/\Omega\right)  ^{2}%
-\lambda_{\mathrm{r}}^{2}}. \label{phi_FRM}%
\end{equation}
Thus the field envelope $\mathbf{u}$ defined in Eq.\thinspace(\ref{u}) is
transformed according to $\mathbf{u}_{\mathrm{out}}\left(  t\right)
=M_{\mathrm{FRM}}\left(  t\right)  \mathbf{u}_{\mathrm{in}}\left(  t\right)  $
with
\begin{equation}
M_{\mathrm{FRM}}\left(  t\right)  =\sqrt{R}\left(
\begin{array}
[c]{cc}%
\cos\left\{  \phi\left[  \Omega\left(  t\right)  \right]  \right\}  &
-\sin\left\{  \phi\left[  \Omega\left(  t\right)  \right]  \right\} \\
\sin\left\{  \phi\left[  \Omega\left(  t\right)  \right]  \right\}  &
\cos\left\{  \phi\left[  \Omega\left(  t\right)  \right]  \right\}
\end{array}
\right)  , \label{MFRM}%
\end{equation}
where the insertion loss of the FRM is captured by the power reflectance $R$.

Here we use a value $\lambda_{\mathrm{r}}=363\,\mathrm{nm}$, which has been
obtained by a fit to experimental data for terbium gallium garnet
\cite{barnes1992variation}. This value also describes well experimental data
for other materials used for FRMs, such as rare-earth iron garnet
\cite{manual_Granopt} as well as terbium-doped glass \cite{TGG}. In the
following, we assume an insertion loss of $1\,\mathrm{dB}$
\cite{presi2011stable}, corresponding to $R\approx0.80$.

\subsection{\label{sec:PC}Polarization Controller}

The polarization controller (PC) consists of three paddles, i.e., fiber coils
with diameters of several cm and multiple windings. These coils are chosen so
that a phase shift between the two polarization components of $\pi/2$, $\pi$
and again $\pi/2$ is obtained, corresponding to a $\lambda/4$, $\lambda/2$ and
$\lambda/4$ waveplate, respectively \cite{wieser2012chromatic}. Each paddle
can be tilted, so that the principal polarization axes are rotated against
each other. In experiment, the PC is manually adjusted until good lasing is
obtained, which can be achieved for different settings
\cite{wieser2012chromatic}. In our model, the PC setting is described by three
orientation angles $\phi_{i}$ ($i=1,2,3$) for the principal polarization axes
of the three coils, where we assume that the associated twisting of the fiber
is a second order effect which can be neglected. Due to the relatively short
fiber length in the PC, we neglect PMD, fiber nonlinearities and dispersion in
Eq.\thinspace(\ref{prop}), only considering the bending birefringence term.
Thus, we obtain for a single coil with fiber length $L_{\mathrm{c}}$ the
solution $\mathbf{u}_{\mathrm{out}}=M\mathbf{u}_{\mathrm{in}}$ with%
\begin{equation}
M\left(  \Omega\right)  =\left(
\begin{array}
[c]{cc}%
\exp\left[  \frac{\mathrm{i}}{2}\Delta_{\beta,\mathrm{b}}\left(
\Omega\right)  L_{\mathrm{c}}\right]  & 0\\
0 & \exp\left[  -\frac{\mathrm{i}}{2}\Delta_{\beta,\mathrm{b}}\left(
\Omega\right)  L_{\mathrm{c}}\right]
\end{array}
\right)  , \label{MWP}%
\end{equation}
where $\Delta_{\beta,\mathrm{b}}\left(  \Omega\right)  $ is given by
Eq.\thinspace(\ref{Dw}). The matrices $M_{\lambda/4}$ and $M_{\lambda/2}$ for
the quarter- and half-waveplate fiber coil are obtained from Eq.\thinspace
(\ref{MWP}) by setting $L_{\mathrm{c}}=\left|  \Delta_{\beta,\mathrm{b}}%
^{-1}\left(  \omega_{\mathrm{c}}\right)  \right|  \pi/2$ and $L_{\mathrm{c}%
}=\left|  \Delta_{\beta,\mathrm{b}}^{-1}\left(  \omega_{\mathrm{c}}\right)
\right|  \pi$, respectively. With the rotation matrix%
\[
M\left(  \phi\right)  =\left(
\begin{array}
[c]{cc}%
\cos\phi & -\sin\phi\\
\sin\phi & \cos\phi
\end{array}
\right)  ,
\]
the total matrix of the polarization controller is obtained as%
\begin{align}
M_{\mathrm{PC}}\left(  t\right)   &  =M\left(  \phi_{3}\right)  M_{\lambda
/4}\left[  \Omega\left(  t\right)  \right]  M\left(  \phi_{2}-\phi_{3}\right)
M_{\lambda/2}\left[  \Omega\left(  t\right)  \right] \nonumber\\
&  \times M\left(  \phi_{1}-\phi_{2}\right)  M_{\lambda/4}\left[
\Omega\left(  t\right)  \right]  M\left(  -\phi_{1}\right)  . \label{MPC}%
\end{align}

\section{Numerical Implementation}

\label{sec:num}The simulation of the FDML dynamics is based on numerically
solving Eq.\thinspace(\ref{prop}). Here, the sweep filter term is evaluated in
Fourier domain, while the other effects are treated in time domain. The
temporal simulation window is adapted to the roundtrip time in the laser
cavity, taking advantage of the implicit periodic boundary conditions of the
numerical scheme \cite{jirauschek2009theoretical,todor2012balance}. The
implementation of the SOA is discussed in Section \ref{sec:FDML}.

The computationally expensive part is computing the optical propagation
through the fiber spool, where the stochastic nature of PMD is reflected by a
randomly varying orientation angle $\theta$ with $\partial_{z}\theta
=g_{\theta}\left(  z\right)  $ as described in Section \ref{sec:theo}. Here,
$g_{\theta}\left(  z\right)  $ is a white noise process with $\left\langle
g_{\theta}\left(  z\right)  \right\rangle =0$ and $\left\langle g_{\theta
}\left(  z\right)  g_{\theta}\left(  z^{\prime}\right)  \right\rangle
=2h_{\mathrm{f}}^{-1}\delta\left(  z-z^{\prime}\right)  $ where $h_{\mathrm{f}%
}$ denotes the fiber autocorrelation length
\cite{wai1994polarization,1996JLwT...14..148W}. This effect is numerically
implemented by dividing the fiber into a large number of small segments of
length $d$, chosen so that $d\ll h_{\mathrm{f}}$. The orientation angle in
each segment $n$ is then randomly chosen using%
\begin{equation}
\theta_{n}=\theta_{0}+\left(  2d/h_{\mathrm{f}}\right)  ^{1/2}\sum_{\ell
=1}^{n}x_{\ell}, \label{thetan}%
\end{equation}
where $\theta_{0}$ is a random variable uniformly distributed on $\left[
0,2\pi\right]  $, and the $x_{\ell}$ are independent random variables
distributed according to the standard normal distribution $f\left(  x_{\ell
}\right)  =\left(  2\pi\right)  ^{-1/2}\exp\left(  -x_{\ell}^{2}/2\right)  $.
For the simulation, the experimental sigma ring geometry shown in
Fig.\thinspace\ref{fig1} is replaced by an equivalent unfolded configuration
with unidirectional propagation through the optical fiber, see Fig.\thinspace
\ref{fig2}. As described in Section \ref{sec:setup}, partial compensation of
the polarization effects in the fiber spool is experimentally achieved by
back-reflecting the light with 90%
${{}^\circ}$
rotated polarization. In our simulation, this is considered by mirror symmetry
of the fiber with respect to the FRM in the unfolded configuration. Thus, in
the simulation the PMD orientation angles $\theta_{n}$ of the fiber segments
$n=1\dots N$ to the left of the FRM in Fig.\thinspace\ref{fig2} are randomly
chosen, and the angles of the mirror segments $n=\left(  N+1\right)  \dots2N$
to the right of the FRM are given by $\theta_{n}=\theta_{2N+1-n}$.
Eq.\thinspace(\ref{prop}) can then be solved numerically based on an
analytical solution for each segment.

For the propagation of optical pulses in a fiber of length $L$, the time delay
$\Delta T$ between the two polarization components due to PMD depends on the
random changes in birefringence, with the variance of $\Delta T$ given by
$\sigma_{\Delta T}^{2}\approx D_{\mathrm{p}}^{2}h_{\mathrm{f}}\left[
\exp\left(  -L/h_{\mathrm{f}}\right)  +L/h_{\mathrm{f}}-1\right]  $
\cite{agr01}. Our PMD implementation based on randomly chosen birefringence
orientation angles, Eq.\thinspace(\ref{thetan}), was validated against this
analytical model for different fiber parameters by propagating Gaussian
optical pulses for many stochastic realizations of $\theta$, and computing
$\sigma_{\Delta T}^{2}$ from the obtained distribution of $\Delta T$.

In an optical fiber, we have $a_{\mathrm{s}}=0$, and typically identical loss
for both polarization directions, $a_{x}=a_{y}=:a$ \cite{agr01}. Furthermore
assuming no fiber gain, $g_{x,y}=0$, the solution of the linear part of
Eq.\thinspace(\ref{prop}), i.e., without SPM\ and XPM, is for segment $n$ at
$z_{n}\leq z<z_{n}+d$ given by%
\begin{align}
\left(
\begin{array}
[c]{c}%
u_{x}\left(  z_{n}+d,t\right) \\
u_{y}\left(  z_{n}+d,t\right)
\end{array}
\right)   &  =\exp\left\{  \mathrm{i}\sum_{j\geq2}\left[  \Omega\left(
t\right)  -\omega_{\mathrm{c}}\right]  ^{j}D_{j}d-ad\right\} \nonumber\\
&  \times M_{\mathrm{L},n}\left(  t\right)  \left(
\begin{array}
[c]{c}%
u_{x}\left(  z_{n},t\right) \\
u_{y}\left(  z_{n},t\right)
\end{array}
\right)  . \label{uL}%
\end{align}
$\Omega\left(  t\right)  $ is determined by the output waveform of the
function generator used to drive the sweep filter. For example, sinusoidal
modulation can be applied, yielding $\Omega\left(  t\right)  =\omega
_{\mathrm{c}}+\left(  \Delta\omega/2\right)  \mathrm{cos}(2\pi t/T_{\mathrm{R}%
})$ with the cavity roundtrip time $T_{\mathrm{R}}$ and sweep range
$\Delta\omega$. The matrix $M_{\mathrm{L},n}$, like the matrices
$M_{\mathrm{FRM}}$ and $M_{\mathrm{PC}}$ defined in Eqs.\thinspace(\ref{MFRM})
and \thinspace(\ref{MPC}), has the form%
\begin{equation}
M=\left(
\begin{array}
[c]{cc}%
A & B\\
-B^{\ast} & A^{\ast}%
\end{array}
\right)  , \label{M}%
\end{equation}
where the asterisk denotes the complex conjugate and the matrix elements are
given by%
\begin{align}
A_{\mathrm{L},n}\left(  t\right)   &  =\cos\left\{  \frac{d}{2}\Delta
_{\beta,n}\left[  \Omega\left(  t\right)  \right]  \right\} \nonumber\\
&  +\mathrm{i}\frac{\Delta_{\beta,\mathrm{b}}\left[  \Omega\left(  t\right)
\right]  +\Delta_{\beta,\mathrm{PMD}}\left[  \Omega\left(  t\right)  \right]
\cos\theta_{n}}{\Delta_{\beta,n}\left[  \Omega\left(  t\right)  \right]  }%
\sin\left\{  \frac{d}{2}\Delta_{\beta,n}\left[  \Omega\left(  t\right)
\right]  \right\}  ,\nonumber\\
B_{\mathrm{L},n}\left(  t\right)   &  =\mathrm{i}\frac{\Delta_{\beta
,\mathrm{PMD}}\left[  \Omega\left(  t\right)  \right]  }{\Delta_{\beta
,n}\left[  \Omega\left(  t\right)  \right]  }\sin\left\{  \frac{d}{2}%
\Delta_{\beta,n}\left[  \Omega\left(  t\right)  \right]  \right\}  \sin
\theta_{n}. \label{ML}%
\end{align}
$\Delta_{\beta,\mathrm{b}}\left(  \Omega\right)  $ and $\Delta_{\beta
,\mathrm{PMD}}\left(  \Omega\right)  $ are defined in Eq.\thinspace(\ref{Dw})
and Eq.\thinspace(\ref{DPMD2}), respectively, and $\Delta_{\beta,n}=\left(
\Delta_{\beta,\mathrm{b}}^{2}+\Delta_{\beta,\mathrm{PMD}}^{2}+2\Delta
_{\beta,\mathrm{b}}\Delta_{\beta,\mathrm{PMD}}\cos\theta_{n}\right)  ^{1/2}$.

The nonlinear part of Eq.\thinspace(\ref{prop}) containing SPM\ and XPM can be
rearranged as%
\begin{equation}
\left.  \partial_{z}\mathbf{u}\right|  _{\mathrm{NL}}=\mathrm{i}\gamma\left[
\frac{5}{6}\left(  \left|  u_{x}^{2}\right|  +\left|  u_{y}^{2}\right|
\right)  +\frac{1}{6}\left(  \left|  u_{x}^{2}\right|  -\left|  u_{y}%
^{2}\right|  \right)  \left(
\begin{array}
[c]{cc}%
1 & 0\\
0 & -1
\end{array}
\right)  \right]  \mathbf{u}.
\end{equation}
The solution is given by%
\begin{align}
\left(
\begin{array}
[c]{c}%
u_{x}\left(  z_{n}+d,t\right) \\
u_{y}\left(  z_{n}+d,t\right)
\end{array}
\right)   &  =\exp\left\{  \mathrm{i}\frac{5}{6}\gamma\left[  \left|
u_{x}^{2}\left(  z_{n},t\right)  \right|  +\left|  u_{y}^{2}\left(
z_{n},t\right)  \right|  \right]  d\right\} \nonumber\\
&  \times M_{\mathrm{NL},n}\left(  t\right)  \left(
\begin{array}
[c]{c}%
u_{x}\left(  z_{n},t\right) \\
u_{y}\left(  z_{n},t\right)
\end{array}
\right)  , \label{uNL}%
\end{align}
where $M_{\mathrm{NL},n}$ again assumes the form Eq.\thinspace(\ref{M}) with
\begin{align}
A_{\mathrm{NL},n}\left(  t\right)   &  =\exp\left\{  \mathrm{i}\frac{\gamma
d}{6}\left[  \left|  u_{x}^{2}\left(  z_{n},t\right)  \right|  -\left|
u_{y}^{2}\left(  z_{n},t\right)  \right|  \right]  \right\}  ,\nonumber\\
B_{\mathrm{NL},n}\left(  t\right)   &  =0. \label{MNL}%
\end{align}
The total matrix for the fiber spool with the FRM can then be obtained by
subsequently propagating the field vector $\mathbf{u}$ through the fiber
segments $n=1\dots N$, the FRM, Eq.\thinspace(\ref{MFRM}), and again the fiber
segments in reverse order. The linear and nonlinear effects are evaluated
separately for each fiber segment using Eqs.\thinspace(\ref{uL}) and
(\ref{uNL}), respectively, where the $\left|  u_{x}^{2}\right|  $ and $\left|
u_{y}^{2}\right|  $ in Eq.\thinspace(\ref{uNL}) have to be updated for each
propagation step.

The fiber loss can in FDML lasers usually be neglected, corresponding to
$a\approx0$. Then the total optical power does not change along the fiber,
i.e., $\left|  u_{x}^{2}\left(  z,t\right)  \right|  +\left|  u_{y}^{2}\left(
z,t\right)  \right|  =\left|  u_{x}^{2}\left(  0,t\right)  \right|  +\left|
u_{y}^{2}\left(  0,t\right)  \right|  =P\left(  t\right)  $ for the fiber
section in Fig.\thinspace\ref{fig2} before the FRM,\ and
\begin{equation}
P_{\mathrm{out}}\left(  t\right)  =R\left(  \left|  u_{x}^{2}\left(
0,t\right)  \right|  +\left|  u_{y}^{2}\left(  0,t\right)  \right|  \right)
=RP\left(  t\right)  \label{Pout}%
\end{equation}
for the section after the FRM. Under this condition, the output field
$\mathbf{u}_{\mathrm{out}}$ of the delay line, consisting of the fiber spool
with length $L$ and FRM, is related to the input field $\mathbf{u}\left(
0\right)  $ by
\begin{align}
\mathbf{u}_{\mathrm{out}}\left(  t\right)   &  =\exp\left\{  \mathrm{i}%
\frac{5}{6}\gamma\left[  \left|  u_{x}^{2}\left(  0,t\right)  \right|
+\left|  u_{y}^{2}\left(  0,t\right)  \right|  \right]  L\left(  1+R\right)
\right\} \nonumber\\
&  \times\exp\left\{  2\mathrm{i}L\sum_{j\geq2}\left[  \Omega\left(  t\right)
-\omega_{\mathrm{c}}\right]  ^{j}D_{j}\right\}  M_{\mathrm{FS}}\left(
t\right)  \mathbf{u}\left(  0,t\right)  , \label{uout}%
\end{align}
with the total matrix%
\begin{equation}
M_{\mathrm{FS}}=M_{2N}M_{2N-1}\dots M_{N+1}M_{\mathrm{FRM}}M_{N}\dots M_{1},
\label{MFS}%
\end{equation}
where $M_{n}=M_{\mathrm{NL},n}M_{\mathrm{L},n}$ with $M_{\mathrm{NL},n}$ and
$M_{\mathrm{L},n}$ given by Eqs.\thinspace(\ref{MNL}) and (\ref{ML}),
respectively. Furthermore, $\theta_{2N+1-n}=\theta_{n}$ in Eq.\thinspace
(\ref{ML}) as discussed above, which results in $M_{\mathrm{L},2N+1-n}%
=M_{\mathrm{L},n}$. The accuracy of this split-step approach can be improved
by adopting a symmetrized procedure \cite{agr01}. Here, the nonlinear effects
are evaluated in the middle of the segments, i.e., the matrix Eq.\thinspace
(\ref{ML}) describing the linear effects is divided into two identical parts
$M_{\mathrm{L},n}\left(  d/2\right)  $ for half the segment width $d/2$,
yielding the matrix $M_{\mathrm{L},n}\left(  d/2\right)  M_{\mathrm{NL}%
,n}\left(  d\right)  M_{\mathrm{L},n}\left(  d/2\right)  $ for segment $n$.

The matrices $M_{\mathrm{FRM}}$, $M_{\mathrm{L},n}$ and $M_{\mathrm{NL},n}$
defined in Eqs.\thinspace(\ref{MFRM}), (\ref{ML}) and (\ref{MNL}), as well as
the matrix representing the polarization controller $M_{\mathrm{PC}}$ given by
Eq.\thinspace(\ref{MPC}), are all of the form Eq.\thinspace(\ref{M}). Products
of such matrices also assume the form Eq.\thinspace(\ref{M}), which can be
exploited to reduce the numerical effort for matrix multiplication since only
the matrix elements in the upper row must be determined.

\subsection{Evaluation of the Fiber Nonlinearity}

\label{sec:non}

Typically the temporal simulation window has to be resolved by several million
grid points for sufficient temporal and spectral resolution. Furthermore, the
optical field is propagated over several $10^{4}$ roundtrips to obtain
convergence. Thus, the fiber propagation matrix $M_{\mathrm{FS}}$,
Eq.\thinspace(\ref{MFS}), has to be precomputed and stored for each sweep
filter frequency value $\Omega\left(  t_{m}\right)  $ associated with a
temporal grid point $t_{m}$. In addition, the dependence of Eq.\thinspace
(\ref{MFS}) on $\left|  u_{x}^{2}\right|  -\left|  u_{y}^{2}\right|  $, which
is introduced by the $M_{\mathrm{NL},n}$\ defined in Eq.\thinspace(\ref{MNL}),
must be considered. This can be done by precomputing $M_{\mathrm{FS}}$ at each
$\Omega\left(  t_{m}\right)  $ for different values $\left|  u_{x}^{2}\right|
-\left|  u_{y}^{2}\right|  $ which are then used for interpolation.

In the following, we derive a suitable interpolation function. As pointed out
above, all the matrices in Eq.\thinspace(\ref{MFS}), and thus also any product
of these matrices, have the form Eq.\thinspace(\ref{M}), so that the $\left|
u_{x,y}^{2}\right|  $ for $M_{\mathrm{NL},n}$ in Eq.\thinspace(\ref{MNL})\ can
be expressed as $\left|  u_{x}\right|  =\left|  Au_{x}\left(  0\right)
+Bu_{y}\left(  0\right)  \right|  $, $\left|  u_{y}\right|  =\left|  -B^{\ast
}u_{x}\left(  0\right)  +A^{\ast}u_{y}\left(  0\right)  \right|  $, where we
have dropped the $t$ dependence for notational convenience. Writing
$u_{x}\left(  0\right)  =P^{1/2}\exp\left(  \mathrm{i}\rho_{x}\right)
\cos\delta$, $u_{y}\left(  0\right)  =P^{1/2}\exp\left(  \mathrm{i}\rho
_{y}\right)  \sin\delta$ with the optical power $P$, phases $\rho_{x,y}$ and
polarization angle $\delta$, we express the phase term in Eq.\thinspace
(\ref{MNL}) as%
\begin{align}
\varphi &  =\frac{\gamma d}{6}\left(  \left|  u_{x}^{2}\right|  -\left|
u_{y}^{2}\right|  \right) \nonumber\\
&  =\frac{\gamma d}{6}P\left\{  \left(  \left|  A\right|  ^{2}-\left|
B\right|  ^{2}\right)  \cos\left(  2\delta\right)  +2\Re\left[  BA^{\ast}%
\exp\left(  \mathrm{i}\rho\right)  \sin\left(  2\delta\right)  \right]
\right\} \nonumber\\
&  =PP_{0}^{-1}\left[  \cos\left(  2\delta\right)  +\kappa\left(  \cos\psi
\cos\rho-\sin\psi\sin\rho\right)  \sin\left(  2\delta\right)  \right]  ,
\label{phi}%
\end{align}
where $P_{0}^{-1}=\gamma d\left(  \left|  A\right|  ^{2}-\left|  B\right|
^{2}\right)  /6$, $\rho=\rho_{y}-\rho_{x}$, $\psi=\angle B-\angle A$, and
$\kappa=2\left|  AB\right|  /\left(  \left|  A\right|  ^{2}-\left|  B\right|
^{2}\right)  $. For the typical parameter values in FDML\ lasers, we have
$\left|  \varphi\right|  \ll1$ and can thus approximate $\exp\left(
\pm\mathrm{i}\varphi\right)  \approx1\pm\mathrm{i}\varphi$. Hence,
Eq.\thinspace(\ref{MNL}) can be written as $M_{\mathrm{NL}}\approx
I+\varphi\Delta_{\mathrm{NL}}$ where the matrix $\Delta_{\mathrm{NL}}$ has the
form Eq.\thinspace(\ref{M}) with $A=\mathrm{i}$, $B=0$, and $I$ denotes the
unit matrix. Consequently, the power dependence of the total propagation
matrix $M_{\mathrm{FS}}$ in Eq.\thinspace(\ref{MFS}) is to first order given
by $M_{\mathrm{FS}}\left(  P\right)  =M_{\mathrm{FS}}\left(  P=0\right)
+\varphi\Delta_{\mathrm{FS}}$, with $\varphi$ given by Eq.\thinspace
(\ref{phi}). The unknowns which have to be determined are $\kappa$, $\psi$ and
the matrix $P_{0}^{-1}\Delta_{\mathrm{FS}}$. Alternatively, we can rewrite
this expression using three dimensionless matrices $\Delta_{1,2,3}$ of the
form Eq.\thinspace(\ref{M}),%
\begin{align}
M_{\mathrm{FS}}\left(  P\right)   &  =M_{\mathrm{FS}}\left(  P=0\right)
+PP_{0}^{-1}\left[  \Delta_{1}\cos\left(  2\delta\right)  \right. \nonumber\\
&  \left.  +\left(  \Delta_{2}\cos\rho+\Delta_{3}\sin\rho\right)  \sin\left(
2\delta\right)  \right]  . \label{Ma}%
\end{align}
In this equation, $P_{0}$ assumes the role of a reference power.
Eq.\thinspace(\ref{Ma}) describes the influences of the fiber nonlinearity
very accurately if $P_{0}$ is set to a typical optical power level in the FDML
fiber spool; here we choose $P_{0}=100\,\mathrm{mW}$. $M_{\mathrm{FS}}\left(
P=0\right)  $ is obtained from Eq.\thinspace(\ref{MFS}) for $u_{x}\left(
0\right)  =u_{y}\left(  0\right)  =0$. The $\Delta_{i}$\ can then be
determined by computing Eq.\thinspace(\ref{MFS}) for three initial fields
$\mathbf{u}\left(  0\right)  $. For example, $u_{x}\left(  0\right)
=P_{0}^{1/2}$, $u_{y}\left(  0\right)  =0$, which corresponds to $\delta=0$
and $P=P_{0}$, yields $\Delta_{1}=\left.  M_{\mathrm{FS}}\left(  P_{0}\right)
\right|  _{\delta=0}-M_{\mathrm{FS}}\left(  P=0\right)  $. Likewise,
$u_{x}\left(  0\right)  =u_{y}\left(  0\right)  =\left(  P_{0}/2\right)
^{1/2}$ yields $\Delta_{2}=\left.  M_{\mathrm{FS}}\left(  P_{0}\right)
\right|  _{\delta=\pi/4,\rho=0}-M_{\mathrm{FS}}\left(  P=0\right)  $, and
$u_{x}\left(  0\right)  =\left(  P_{0}/2\right)  ^{1/2}$, $u_{y}\left(
0\right)  =\mathrm{i}\left(  P_{0}/2\right)  ^{1/2}$ gives $\Delta_{3}=\left.
M_{\mathrm{FS}}\left(  P_{0}\right)  \right|  _{\delta=\pi/4,\rho=\pi
/2}-M_{\mathrm{FS}}\left(  P=0\right)  $. The accuracy of this propagation
matrix can be further improved by strictly enforcing the relation between
output and input power given in Eq.\thinspace(\ref{Pout}). This yields the
improved matrix%
\begin{equation}
M_{\mathrm{FS,corr}}\left(  P\right)  =\left[  R/\left|  M_{\mathrm{FS}%
}\left(  P\right)  \right|  \right]  ^{1/2}M_{\mathrm{FS}}\left(  P\right)  ,
\label{Ma2}%
\end{equation}
where $\left|  M_{\mathrm{FS}}\left(  P\right)  \right|  $ denotes the matrix determinant.

\section{Results}

\label{sec:results}

In the following, we use the numerical approach developed in Section
\ref{sec:num} to simulate propagation through a fiber spool and operation of
an FDML laser, respectively. Fiber bending birefringence and PMD are
implemented as described in Sections \ref{sec:BB} and \ref{sec:PMD},
respectively, and the nonlinearity is treated according to Eqs.\thinspace
(\ref{Ma}) and \thinspace(\ref{Ma2}). The FRM is implemented as described in
Section \ref{sec:FRM}. In our simulations, we use a fiber segment length of
$d=0.005\,\mathrm{m}$ to obtain sufficient spatial resolution.

\subsection{Propagation through a Fiber Spool}

In the following, we simulate the propagation of wavelength-swept, linearly
polarized light through a fiber spool, and compare the obtained output state
of polarization (SOP) to experimental results \cite{wieser2012chromatic}. We
study propagation through a $1\,\mathrm{km}$ spool with a bending radius
$R_{\mathrm{b}}=0.115\,\mathrm{m}$ \cite{wieser2012chromatic}, where the light
is back-reflected at the fiber end using an FRM to obtain a partial
compensation of polarization effects and a doubling of the propagation length.
Such a system is employed in FDML\ lasers as a delay line, see Fig.\thinspace
\ref{fig1}. We use the fiber dispersion given in the manufacturer
specifications \cite{SMF28}, and assume a nonlinear coefficient of
$\gamma=0.0011\,\mathrm{W}^{-1}\mathrm{m}^{-1}$. For comparison, also
propagation through a traditional $2\,\mathrm{km}$ fiber spool is considered
where polarization effects are not compensated for.

The SOP of an optical field $\mathbf{u}=\left[  u_{x},u_{y}\right]
^{\mathrm{T}}$ can be described by the Stokes vector components $S_{0}=\left|
u_{x}^{2}\right|  +\left|  u_{y}^{2}\right|  $, $S_{1}=\left|  u_{x}%
^{2}\right|  -\left|  u_{y}^{2}\right|  $, $S_{2}=2\Re(u_{x}u_{y}^{\ast})$,
$S_{3}=2\Im(u_{x}u_{y}^{\ast})$, and the Poincar\'{e} sphere representation is
in spherical coordinates given by \cite{wieser2012chromatic}
\begin{align*}
p  &  =\left(  S_{1}^{2}+S_{2}^{2}+S_{3}^{2}\right)  ^{1/2}S_{0}^{-1},\\
\Theta &  =\angle(S_{1}+\mathrm{i}S_{2}),\\
\Phi &  =\mathrm{asin}\left[  S_{3}\left(  S_{1}^{2}+S_{2}^{2}+S_{3}%
^{2}\right)  ^{-1/2}\right]  .
\end{align*}
In our simulation we assume fully polarized light, i.e., the polarization
degree is $p=1$, and thus the polarization state is completely characterized
by the two angles: $\Theta$ describes the linear polarization direction, with
$\Theta/2$ being the orientation angle of the field. $\Phi$ characterizes the
circularity, with $\pm90%
{{}^\circ}%
$ denoting left and right circular polarization, respectively.

\begin{figure}[ptbh]
\centerline{\includegraphics[width=8.4cm]{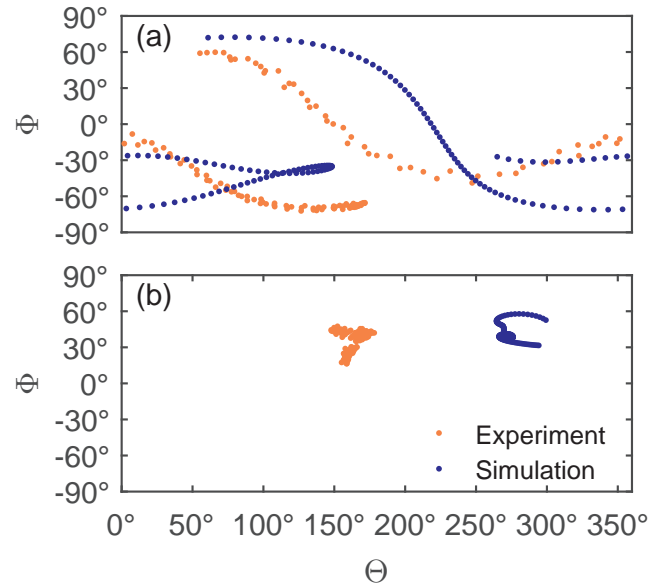}}\caption{Simulated and
experimental SOP after $2\,\mathrm{km}$ propagation through an (a)
uncompensated and (b) compensated fiber\ spool. The input field is
wavelength-swept between $1246-1382\,\mathrm{nm}$ and is linearly polarized at
$45{{}^{\circ}}$ to the spool plane.}%
\label{fig3}%
\end{figure}

In our simulation, the stochastic nature of PMD is reflected by a random
choice of the birefringence orientation angle $\theta$ according to
Eq.\thinspace(\ref{thetan}). Thus, the obtained results greatly depend on the
exact realization of $\theta$. In Fig.\thinspace\ref{fig3}, simulated and
experimental SOP traces are shown for a wavelength sweep of
$1246-1382\,\mathrm{nm}$, with the polarization axis of the incoming field
oriented at an angle of $45%
{{}^\circ}%
$ to the spool plane. In Fig.\thinspace\ref{fig3}(a), data for the traditional
$2\,\mathrm{km}$ fiber spool are shown. Fig.\thinspace\ref{fig3}(b) contains
results for the fiber system with the FRM included, consisting of the
compensated $1\,\mathrm{km}$ spool and additionally two uncompensated
$2\,\mathrm{m}$ spools (bending radius $3.5\,\mathrm{cm}$) in front and behind
it \cite{wieser2012chromatic}. As expected, the experimental and theoretical
traces do not completely coincide due to the random PMD contribution, and also
because the experimental measurement setup contains some additional components
not considered in the simulation, such as a circulator for the compensated
spool. Furthermore, for the bending radii considered, small variations in the
length of non-compensated fiber stretches on the order of $1\,\mathrm{m}$ can
already significantly alter the position of the output SOP trajectory on the
Poincar\'{e} sphere \cite{wieser2012chromatic}. Thus, the only feature of the
simulated and measured SOP trajectories that can be expected to be comparable
is their extent in the $\Theta\Phi$-plane. Although the obtained trajectories
will be different for each realization of $\theta$, their approximate extent
is related to the PMD parameter $D_{\mathrm{p}}$, characterizing the PMD
strength and thus the mean PMD-induced deviation of the output SOP from the
input SOP. We find that the extent of the computed and measured trajectories
in both $\Theta$ and $\Phi$ is similar, confirming that the influence of
birefringence on the polarization state dynamics is of comparable strength in
simulation and experiment: For the uncompensated case shown in Fig.\thinspace
\ref{fig3}(a), the angular spread $\Delta\Phi\approx140%
{{}^\circ}%
$, and $\Delta\Theta$ even exceeds $360%
{{}^\circ}%
$. The spread is much smaller for the polarization compensated spool in
Fig.\thinspace\ref{fig3}(b), with $\Delta\Theta\approx$ $\Delta\Phi\approx30%
{{}^\circ}%
$, demonstrating the effectiveness of the polarization compensation scheme.

\begin{figure}[ptbh]
\centerline{\includegraphics[width=8.4cm]{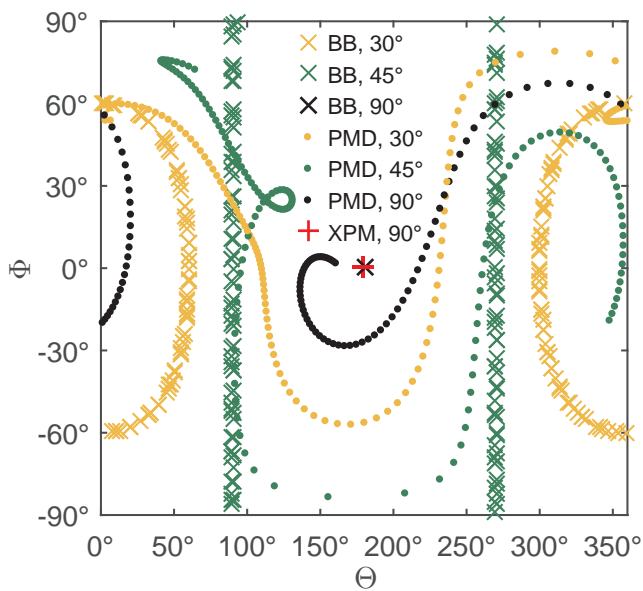}}\caption{Simulated SOP
after propagation of a wavelength-swept ($1246-1382\,\mathrm{nm}$)\ field
through an uncompensated fiber\ spool for different input polarization angles,
taking into account only polarization effects due to bending birefringence
(BB), PMD or XPM/SPM, respectively.}%
\label{fig4}%
\end{figure}

In Fig.\thinspace\ref{fig4}, the simulated SOP trace for the
uncompensated\ spool is shown, taking into account only polarization effects
due to bending birefringence, PMD or XPM/SPM, respectively. For bending
birefringence, the fiber spool acts as a waveplate, generally converting the
linear input SOP into some elliptical SOP, as shown for a $30%
{{}^\circ}%
$ input polarization angle. The resulting SOP\ trajctory is traversed several
times during a wavelength sweep from $1246-1382\,\mathrm{nm}$, corresponding
to a change in waveplate order. For a $45%
{{}^\circ}%
$ input polarization angle, we obtain rotated linear polarization states for
all wavelengths, while for $0%
{{}^\circ}%
$ (not shown in the figure) and $90%
{{}^\circ}%
$ the linear input SOP remains unchanged. Also PMD converts the linear input
SOP into a generally elliptical SOP. In contrast to bending birefringence, the
SOP\ trajctory is not closed, and is thus not traversed several times during a
wavelength sweep. For the chosen parameters and input angles which do not
approach $0%
{{}^\circ}%
$ or $90%
{{}^\circ}%
$, the angular extent of the SOP traces is comparable for bending
birefringence and PMD. By contrast, as exemplarily shown in Fig.\thinspace
\ref{fig4} for an input polarization angle of $90%
{{}^\circ}%
$, the nonlinear effects due to XPM and SPM do not significantly alter the
input SOP even for an optical power of $1\,\mathrm{W}$\ which already exceeds
typical power levels in FDML fiber spools \cite{mao2009high}.

\begin{figure}[ptbh]
\centerline{\includegraphics[width=8.4cm]{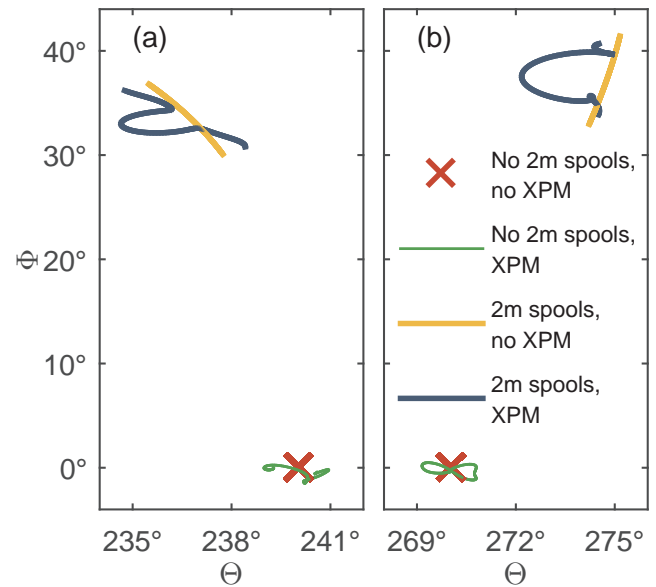}}\caption{Simulated SOP
after propagation of a wavelength-swept ($1246-1382\,\mathrm{nm}$)\ field
through the compensated fiber\ spool for an input polarization angle of (a)
$30{{}^{\circ}}$ and (b) $45{{}^{\circ}}$. Here, an ideal wavelength
independent FRM is assumed. Results are shown with and without including the
two uncompensated $2\,\mathrm{m}$ spools and and fiber nonlinearities due to
XPM/SPM.}%
\label{fig5}%
\end{figure}

In Fig.\thinspace\ref{fig5}, the simulated SOP trace for the compensated spool
is shown, assuming ideal polarization compensation by a wavelength independent
FRM. Residual polarization effects are then only due to fiber nonlinearity,
i.e., XPM and SPM, as well as further, uncompensated components such as the
additional $2\,\mathrm{m}$ spools in the measurement setup. While for ideal
compensation, the input and output SOPs are identical, the fiber
nonlinearity\ breaks the symmetry between forward and backward propagation,
impeding perfect compensation of polarization effects. For an optical power of
$1\,\mathrm{W}$\ assumed here, the deviation of the output from the input SOP
does however not exceed $1.5%
{{}^\circ}%
$ in both $\Theta$\ and $\Phi$. As expected, the uncompensated $2\,\mathrm{m}$
spools alter the position of the output SOP on the Poincar\'{e} sphere, and
lead to an additional angular spread of the SOP trace. By comparison of
Fig.\thinspace\ref{fig5} with Fig.\thinspace\ref{fig3}(b), we can see that the
angular spread of the SOP traces due fiber nonlinearities or additional
uncompensated $2\,\mathrm{m}$ spools is however much smaller than the spread
due to a non-ideal FRM. This demonstrates that residual polarization effects
in the compensated spool are mainly due to the wavelength dependence of the
FRM as modeled by Eq.\thinspace(\ref{phi_FRM}), allowing optimum compensation
only at the FRM design wavelength $\lambda_{\mathrm{F}}$.

\subsection{\label{sec:FDML}FDML\ Laser}

For FDML simulations, where the optical field is propagated over many
roundtrips, the evaluation of Eq.\thinspace(\ref{prop}) using a conventional
split-step approach without precomputed matrices is computationally much
slower than the numerical method developed in Section \ref{sec:num} where the
matrices $M_{\mathrm{FS}}\left(  P=0\right)  $ and $\Delta_{1,2,3}$ in
Eq.\thinspace(\ref{Ma}) have to be calculated only once at the beginning of
the simulation. In this case, only the numerical load for the initial
computation of these matrices increases for a finer spatial discretization,
i.e., a higher number $2N$ of discrete fiber segments used\ (see
Fig.\thinspace\ref{fig2}). By contrast, for the conventional split-step
implementation, the computation time scales with $2N$ for every roundtrip. For
a total fiber propagation distance of $2\times262\,\mathrm{m}=524\,\mathrm{m}%
$, we obtain $2N=104800$ segments with a length of $d=0.005\,\mathrm{m}$. In
this case, we found the numerical propagation through the fiber spool with the
conventional split-step method to be about $30000$ times slower than for the
matrix approach with precomputed matrices. The treatment of the lumped optical
elements, such as the SOA and the sweep filter, is identical in both
approaches. The evaluation of the fiber propagation with precomputed matrices
consumes only $20\%$ of the computational time per roundtrip, while $80\%$ is
required for the other optical resonator elements. Consequently, for the
conventional split-step method the total simulation time increases by a factor
of about $30000\times0.2=6000$, which is prohibitive for the large number of
roundtrips required for convergence.

\begin{figure}[ptbh]
\centerline{\includegraphics[width=8.4cm]{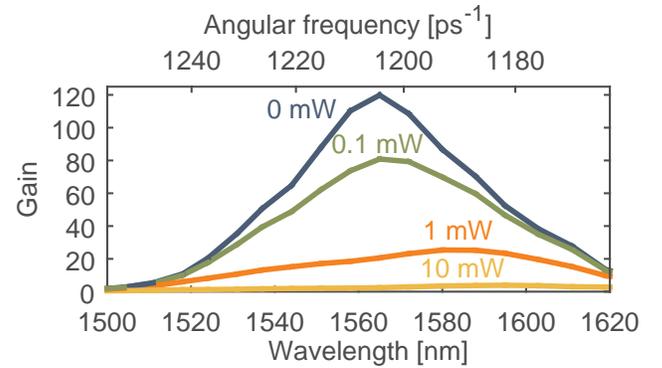}}\caption{SOA power gain
$G_{x}^{2}$ as a function of wavelength and angular frequency for various
values of the incident optical power.}%
\label{figgain}%
\end{figure}

In the following, we present simulation results for an FDML laser used for the
generation of picosecond pulses \cite{eigenwillig2013picosecond}. The setup,
illustrated in Fig.\thinspace\ref{fig1}, is the $4\times$ pass configuration
of Ref.\thinspace\cite{eigenwillig2013picosecond}, which is particularly
interesting because it exhibits pronounced polarization effects. Here, the
sweep filter center frequency is given by $\Omega\left(  t\right)
=\omega_{\mathrm{c}}+\left(  \Delta\omega/2\right)  \mathrm{cos}(2\pi
t/T_{\mathrm{R}})$, where $\omega_{\mathrm{c}}=1.207\times10^{15}%
\,\mathrm{s}^{-1}$ and $\Delta\omega=4.083\times10^{13}\,\mathrm{s}^{-1}$,
corresponding to a central wavelength of $1560\,\mathrm{nm}$ and a sweep
bandwidth of $52.75\,\mathrm{nm}$. The cavity roundtrip time is $T_{\mathrm{R}%
}=2.561\,\mu\mathrm{s}$. The fiber spool ($R_{\mathrm{b}}=0.125\,\mathrm{m}$)
consists of $246\,\mathrm{m}$ single-mode fiber combined with $16\,\mathrm{m}$
of dispersion compensation fiber (DCF). The chromatic dispersion is modeled
based on manufacturer specifications \cite{SMF28} and experimental
characterization data \cite{wieser2009ultra}, respectively, and a nonlinear
coefficient of $\gamma=0.0011\,\mathrm{W}^{-1}\mathrm{m}^{-1}$ is assumed. The
setup features a polarization insensitive SOA (Covega SOA1117), which is
implemented as a lumped element with a frequency-dependent amplitude gain
$G_{x,y}\left(  \omega\right)  =\exp\left[  \int g_{x,y}\left(  \omega\right)
\mathrm{d}z\right]  $, obtained by integrating over the gain medium length.
Additionally, we account for saturation effects by introducing a gain
dependence on the input optical power $P_{\mathrm{in}}$ as shown in
Fig.\thinspace\ref{figgain}, modeled according to experimental
characterization data. The non-instantaneous gain recovery dynamics can be
considered by using a time-averaged optical power for determining the
saturated gain value, e.g., in the limit of slow recovery, by averaging
$P_{\mathrm{in}}$ over a cavity roundtrip \cite{wang2013comparison}. Here we
have to modify this approach since the gain recovery time $T_{\mathrm{L}%
}\approx380\,\mathrm{ps}^{-1}$ of the SOA in the investigated setup is
considerably shorter than the roundtrip time, by using an exponential moving
average value $\bar{P}_{\mathrm{in}}\left(  t\right)  =T_{\mathrm{L}}^{-1}%
\int_{-\infty}^{t}P_{\mathrm{in}}\left(  \tau\right)  \exp\left[  \left(
\tau-t\right)  /T_{\mathrm{L}}\right]  \mathrm{d}\tau$
\cite{jirauschek2009theoretical,todor2012balance}. On a discrete temporal grid
with points $t_{m}$ and spacing $\Delta_{t}$, $\bar{P}_{\mathrm{in}}\left(
t_{m}\right)  $ can then be updated in an efficient way,
\begin{equation}
\bar{P}_{\mathrm{in}}\left(  t_{m}\right)  =\bar{P}_{\mathrm{in}}\left(
t_{m-1}\right)  \exp\left(  -\frac{\Delta_{t}}{T_{\mathrm{L}}}\right)
+\left[  1-\exp\left(  -\frac{\Delta_{t}}{T_{\mathrm{L}}}\right)  \right]
P_{\mathrm{in}}\left(  t_{m}\right)  .
\end{equation}
ASE is implemented as described in Section \ref{sec:theo}, adding noise and
allowing the simulation to self-start
\cite{jirauschek2009theoretical,todor2012balance}. Furthermore, a weakly
polarization dependent gain of $2\,\mathrm{dB}$ of the SOA module is
considered in our model \cite{wieser2012chromatic}. The ring cavity of the
experimental setup contains a polarization controller, here modeled as
described in Section \ref{sec:PC}, which is in principle not required for
designs with polarization insensitive SOA but can bring some performance
improvement \cite{wieser2012chromatic}. Temporal compression of the outcoupled
wavelength-swept FDML light is achieved by negative dispersion, experimentally
provided by a $15\,\mathrm{km}$ DCF which is passed four times in the
$4\times$ pass configuration. Here, only $1.5\,\mathrm{nm}$ sweep range,
corresponding to a sweep duration of $24\,\mathrm{ns}$, is used for pulse
compression in order to reduce higher order dispersion effects in the
DCF\thinspace\cite{eigenwillig2013picosecond}.

To ensure convergence to steady state, the pulse propagation is simulated over
$500000$ roundtrips, corresponding to an evolution of the optical field over
$1.3\,\mathrm{s}$. In analogy to the experiment, the pulse intensity profile
is numerically evaluated by averaging over multiple roundtrips, which leads to
a suppression of fluctuations. In our simulation, we take every $1000$th
roundtrip of the final $20000$ roundtrips. Furthermore, we fine-tune the
length of the fiber compressor so as to obtain optimum compression. To emulate
the experimental conditions where the pulse width is extracted assuming a
Gaussian shape\thinspace\cite{eigenwillig2013picosecond}, we determine the
full width at half-maximum (FWHM) duration by performing a least-square
Gaussian pulse fit to the simulated intensity profile. Similar results are
obtained for the $x$ and $y$ components of the electric field; the simulation
data presented in the following are for the total field, i.e., including both components.

\begin{figure}[ptbh]
\centerline{\includegraphics[width=8.4cm]{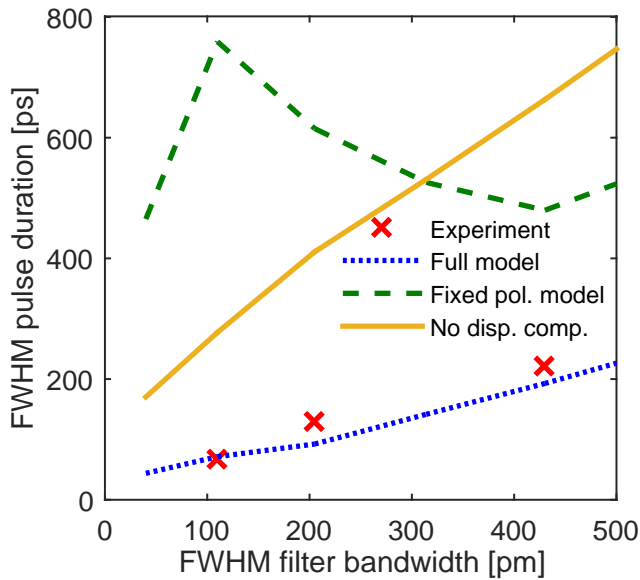}}\caption{Comparison of
experimental and simulated pulse duration for the $4\times$ pass configuration
as a function of the sweep filter bandwidth. Shown are experimental data
\cite{eigenwillig2013picosecond}, along with simulation results obtained by
the full simulation approach of Section \ref{sec:num} and the fixed
polarization model developed for PM FDML
lasers\ \cite{jirauschek2009theoretical}. Additionally, full simulation data
for a laser setup without dispersion compensation is displayed.}%
\label{fig6}%
\end{figure}

In Fig.\thinspace\ref{fig6}, a comparison of the experimental and simulated
FWHM pulse widths for the $4\times$ pass configuration of Ref.\thinspace
\cite{eigenwillig2013picosecond} is shown as a function of the sweep filter
bandwidth. As expected, a narrower filter bandwidth results in a shorter
pulse, since the linewidth is reduced and thus the coherence length improves.
In experiment, the filter bandwith can however not be reduced below
$110\,\mathrm{pm}$ due to mechanical restrictions. Here, the shortest pulse
width of $68\,\mathrm{ps}$ is measured, which agrees very well with the
numerical value of $71.4\,\mathrm{ps}$ obtained by the full simulation
approach of Section \ref{sec:num}. Very good agreement between experiment and
theory is also found for the other sweep filter bandwidths investigated. In
contrast, the model assuming fixed polarization as obtained for a PM
configuration\ \cite{jirauschek2009theoretical} yields much longer pulse
durations, e.g., $759.2\,\mathrm{ps}$ at a filter bandwidth of
$110\,\mathrm{pm}$, exceeding the measured value by a factor of more than
$10$. Experimentally, the instantaneous linewidth (averaged over a sweep) is
estimated to be around\ $10\,\mathrm{pm}$, which is more than a factor of $10$
below the filter bandwidth. Due to the very high sweep speed, an indirect
measurement technique is used, based on evaluating the interference signal
from a Michelson interferometer\thinspace\cite{eigenwillig2013picosecond}. By
contrast, in the simulation the instantaneous linewidth can directly be
determined by Fourier transforming the field envelope defined in
Eq.\thinspace(\ref{u}), since here the frequency axis moves along with the
sweep filter center frequency and smearing of the lineshape due to the sweep
filter dynamics is thus avoided \cite{todor2011instantaneous,todor2012balance}%
. An FWHM instantaneous linewidth of $7.44\,\mathrm{pm}$ is obtained for the
full simulation in reasonable agreement with the experimental estimate, given
the uncertainty of the measurement. The fixed polarization model yields
$46.8\,\mathrm{pm}$ linewidth, implying a much lower coherence and
substantially reduced compressibility of the optical field, and thus resulting
in much longer pulses than obtained by the experiment and full simulation.
This demonstrates that FDML operation is significantly affected by
polarization effects,\ leading to considerably improved coherence for setups
with polarization insensitive SOA where the polarization dynamics can freely
unfold. More detailed simulations show that over a wide parameter range, the
obtained pulse duration and linewidth only weakly depend on the exact value of
bending birefringence, PMD and XPM as quantified by $\Delta_{\beta,\mathrm{b}%
}$, $\Delta_{\beta,\mathrm{PMD}}$ and $\gamma$ in Eq.\thinspace(\ref{prop}).
Also assuming an ideal FRM, as done in Fig.\thinspace\ref{fig5}, does not
change the simulation results significantly. By contrast, the amount of
residual intracavity dispersion plays an important role, and thorough
dispersion compensation is crucial to obtain short pulses. For illustration,
simulation results for an uncompensated FDML setup are included in
Fig.\thinspace\ref{fig6}, where the $16\,\mathrm{m}$ DCF is replaced by a
normal SMF. The obtained pulse durations, as well as the corresponding
linewidths, are larger by a factor of $\sim3-4$ as compared to the full
simulation results for the dispersion compensated setup.

\begin{figure}[ptbh]
\centerline{\includegraphics[width=8.4cm]{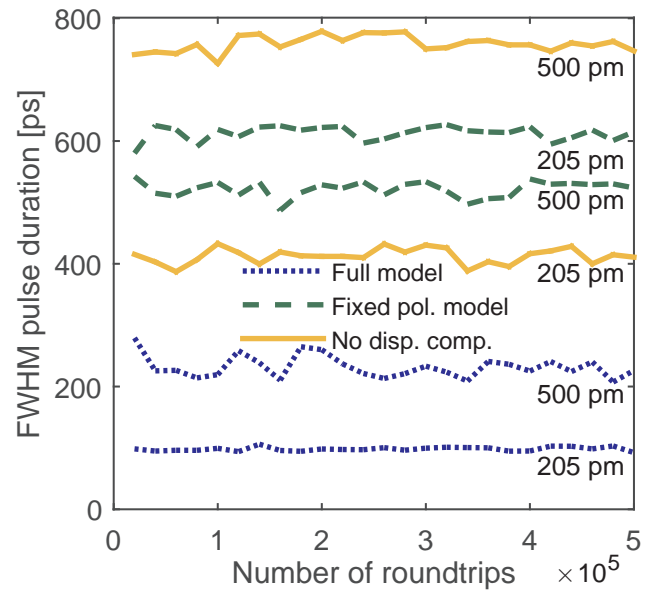}}\caption{Simulated pulse
duration for the $4\times$ pass configuration as a function of the roundtrip
number for sweep filter bandwidths of $205\,\mathrm{pm}$ and $500\,\mathrm{pm}%
$, respectively.}%
\label{fig7}%
\end{figure}To ensure convergence of the simulation results of Fig.\thinspace
\ref{fig6}, the computed pulse duration is displayed in Fig.\thinspace
\ref{fig7} as a function of the roundtrip number for the full simulation
approach with and without dispersion compensation, as well as for the fixed
polarization model. Again, for each datapoint we average over $20$ pulses,
considering every $1000$th of $20000$ successive roundtrips. Apart from small
variations due to the sweep-to-sweep jitter \cite{biedermann2010direct}, the
pulse duration does not change significantly anymore after a few $100000$
roundtrips or less, depending on the laser parameters.

All in all, the simulation results of Fig.\thinspace\ref{fig6} substantiate
experimental observations that low residual intracavity dispersion is crucial
for high-coherence operation of FDML lasers
\cite{biedermann2009dispersion,eigenwillig2013picosecond}, and that the
smallest linewidths down to a few pm are obtained for non-PM configurations
with polarization insensitive SOA.

\section{Conclusion}

\label{sec:conclusion}

In conclusion, the paper provides a theoretical model and its efficient
numerical implementation for the simulation of wavelength-swept waveform
propagation in optical fiber systems, including the polarization dynamics in
fiber spools and other polarization dependent optical components. We
demonstrate that this approach allows for a realistic simulation of non-PM
FDML operation over several $10^{5}$ roundtrips, yielding good agreement with
experimental data. The simulation approach is used to investigate the
influence of polarization effects such as bending birefringence, PMD and XPM
on the wavelength-swept waveform propagation in fiber spools and on FDML laser
operation. The simulation results substantiate experimental observations that
high-coherence operation of FDML lasers requires low residual intracavity
dispersion, and that the smallest linewidths down to a few pm are obtained for
non-PM configurations with polarization insensitive SOA. The developed
numerical model can also be of general relevance for tunable lasers which use
external fiber delay spools to increase the wavelength sweep speed in
applications such as OCT \cite{hong2015optically,potsaid2010ultrahigh}.

\bigskip

\textbf{Funding.} Deutsche Forschungsgemeinschaft (DFG) (JI 115/4-1, JI
115/8-1, HU 1006/6); European Union project ENCOMOLE-2i (Horizon 2020, ERC CoG
no. 646669).

\bigskip


\ifthenelse{\equal{\journalref}{ol}}{\clearpage\bibliographyfullrefs{sample} }{}


\ifthenelse{\equal{\journalref}{aop}}{\section*{Author Biographies}
\begingroup\setlength\intextsep{0pt}
\begin{minipage}
[t][6.3cm][t]{1.0\textwidth}
\begin{wrapfigure}
{L}{0.25\textwidth}     \includegraphics[width=0.25\textwidth
]{john_smith.eps}
\end{wrapfigure}
\noindent{\bfseries John Smith} received his BSc (Mathematics) in 2000 from
The University of Maryland. His research interests include lasers and optics.
\end{minipage}
\begin{minipage}
{1.0\textwidth}
\begin{wrapfigure}
{L}{0.25\textwidth}     \includegraphics[width=0.25\textwidth
]{alice_smith.eps}
\end{wrapfigure}
\noindent{\bfseries Alice Smith} also received her BSc (Mathematics) in 2000
from The University of Maryland. Her research interests also include lasers
and optics.
\end{minipage}
\endgroup}{}
\end{document}